\documentclass[a4paper,floatfix,superscriptaddress,pra,twocolumn,longbibliography]{revtex4-1}

\usepackage{amsmath, amsthm, amssymb}
\usepackage{graphicx}
\usepackage{dcolumn}
\usepackage{bm}
\usepackage{bbm}
\usepackage{color}

\usepackage[usenames,dvipsnames]{xcolor}
\usepackage{svg}
\usepackage{hyperref}
\usepackage{enumerate}
\usepackage{ulem}
\usepackage{textcomp}
\usepackage{blkarray}
\usepackage{bigstrut}
\usepackage{array}
\usepackage{multirow}

\newcommand{\ba}{\begin{eqnarray}}
\newcommand{\ea}{\end{eqnarray}}

\usepackage{float}
\usepackage{placeins}

\begin{document}

\title{Evolution of cooperation in networks with well-connected cooperators}

\author{Josefine Bohr Brask} \thanks{jbb@sodas.ku.dk} \affiliation{SODAS (Copenhagen Center for Social Data Science), University of Copenhagen, Copenhagen, Denmark} \affiliation{Section for Ecology and Evolution, Department of Biology, University of Copenhagen, Copenhagen, Denmark} \affiliation{Department of Applied Mathematics and Computer Science (DTU Compute), Technical University of Denmark, Kongens Lyngby, Denmark}
\author{Jonatan Bohr Brask}\affiliation{Center for Macroscopic Quantum States (bigQ), Department of Physics, Technical University of Denmark, Kongens Lyngby, Denmark}

\begin{abstract}
Cooperative behavior constitutes a key aspect of human society and non-human animal systems, but explaining how cooperation evolves represents a major scientific challenge. It is now well established that social network structure plays a central role for the viability of cooperation. However, not much is known about the importance of the positions of cooperators in the networks for the evolution of cooperation. Here, we investigate how the spread of cooperation is affected by correlations between cooperativeness and individual social connectedness (such that cooperators occupy well-connected network positions). Using simulation models, we find that these correlations enhance cooperation in standard scale-free networks but not in standard Poisson networks. In contrast, when degree assortativity is increased such that individuals cluster with others of similar social connectedness, we find that Poisson networks can maintain high levels of cooperation, which can even exceed those of scale-free networks. We show that this is due to dynamics where bridge areas between social clusters act as barriers to the spread of defection. We also find that this positive effect on cooperation is sensitive to the presence of Trojan horses (defectors placed within cooperator clusters), which allow defection to invade. The results provide new knowledge about the conditions under which cooperation may evolve, and are also relevant to consider in regard to the design of cooperation studies.\\


\end{abstract}

\maketitle

\section{Introduction}

Cooperation, understood as behavior where individuals help others, has long constituted a major conundrum for science. Cooperative behavior seemingly contradicts the central prediction of Darwinian evolutionary theory that individuals will behave in ways that maximize their own gain. Nevertheless, cooperation is found widely across species, including humans, other mammals, fish, birds, insects, and microscopic organisms (\cite{brask2019evolution,brucks2020parrots,Carter2013, Queller1998, rand2013human, Rutte2007, Turner1999, Voelkl2015}). Explaining the evolution of cooperation has been called one of the biggest scientific challenges of our time (\cite{Pennisi2005}). 

During recent decades, it has become clear that social network structure plays an essential role for the evolution of cooperation (\cite{gokcekus2021exploring,jusup2022social, Roca2009,Szabo2007}). Simulation studies have shown that while unstructured populations are not conducive to cooperation, certain network structures can promote and stabilize it (e.g. \cite{Assenza2008, Nowak1992,Ohtsuki2006,Santos2005,Santos2006evol}). There is also now extensive empirical evidence that nonrandom social network structure is ubiquitously present in different species (\cite{brask2021,Krause2015}). This implies that the study of cooperation in social structures is of general importance for understanding the evolution cooperative behavior. 

A social network can be described as a set of nodes and a set of edges, where each node constitutes an individual and the edges constitute their social connections. A large body of work has investigated how the spread and viability of cooperation is affected by different structural network features, such as degree distribution (\cite{Santos2005,Santos2006}), degree heterogeneity (\cite{Santos2006evol,Santos2012}), average degree (\cite{Ohtsuki2006}), degree assortativity (\cite{duh2019assortativity, Rong2007, wang2014degree}), clustering (\cite{Assenza2008, david2023clustering}), and modularity (\cite{gianetto2015network, Voelkl2009}; see \cite{jusup2022social, Roca2009, Szabo2007} for reviews). However, while the role of network structure in the evolution of cooperation thus has been extensively investigated, little is known about the role of the positions of cooperative and defective individuals in the network (in other words, the role of correlations between cooperative strategy and node properties such as connectedness; \cite{Chen2008}). Across species, real-world social networks are characterized by considerable heterogeneity in the social positions of individuals, where some are more socially connected than others (\cite{brask2021,Krause2015}). If cooperative individuals tend to have specific social network positions, this may affect the spread and persistence of cooperation. In particular, it may intuitively be expected that cooperation should fare better in situations where cooperative individuals have more social connections (higher degree) than defective individuals, as their higher connectedness potentially could help them propagate their strategy. Such effects could play an important role for the evolution of cooperation, but they are currently not well understood. 

Here, we investigate how correlations between cooperative strategy and social connectedness affect the evolution of cooperation in different network structures. This can elucidate the importance of the network positions of cooperative individuals, and thereby increase our general understanding of the conditions under which cooperation can persist. The investigations are also relevant for situations where social networks are deliberately constructed. For example, in experiments with humans playing cooperation games in artificial network structures (e.g. \cite{Cassar2007, Gracia2012, grujic2020people, li2018punishment, melamed2018cooperation, Rand2014}), initial stochastic correlations between cooperativeness and network position could potentially have a significant effect on the results and influence the conclusions of the experiments, in particular because the number of replications in such experiments can be low due to practical constraints. 

We use a standard methodological approach for investigating the evolution of cooperation in networks, namely game theory-based simulation modeling. Evolutionary game theory provides a common framework for the study of cooperation (\cite{leimar2023game, Smith1973, Smith1982, traulsen2023future}). In this approach, the interaction between individuals is formalized as games where each player adopts one of a limited number of strategies, such as  \textit{cooperative} or \textit{defective} (selfish). The game is played repeatedly and players adapt their strategies to optimize their performance in terms of game  payoffs. Alternatively, each iteration can be interpreted as a reproductive generation, so that the adaptation is genetic. With this framework, it becomes possible to study the stability and dynamics of cooperation by looking at temporal changes in the frequency of cooperators. The evolution of cooperation in structured populations can be studied by simulating games on network structures where interactions occur over the network edges (the social connections), which has been done widely (reviewed in \cite{jusup2022social, Roca2009, Szabo2007}). A key outcome of the simulations is the fraction of cooperators in the population after a large number of timesteps (game rounds), which indicates the long-term viability of cooperation in the network. The simulations most often
begin with the strategies being randomly assigned to the network nodes. In other words, usually cooperators and defectors are initially randomly positioned in the network (\cite{Roca2009, Szabo2007}). 
\vspace{\baselineskip}

In this study, we investigate the effect of nonrandom strategy positions on the spread and survival of cooperation. We specifically study the effect of correlations between cooperative strategy and node degree. We focus on degree because this is a fundamental measure of network position that is easily interpretable (as the number of social interaction partners an individual has). Our approach is to run simulations of the evolution of cooperation in networks, where cooperators are initially placed on well-connected (high-degree) nodes (stochastically or deterministically), and compare the results to those of corresponding simulations where the cooperators are placed randomly in the networks. We note that the simulations can be interpreted both as: (1) simulating systems where the correlations between strategy and network position is a result of the networks being deliberately constructed, for example as in human cooperation experiments; and (2) simulating systems where the correlations between strategy and network position is a result of previous evolution. Thus, the positioning of cooperators on well-connected nodes can be viewed either as the system’s actual initial conditions, or as a transient state. In either case, the simulations show how cooperators’ occupation of well-connected nodes affects subsequent strategy evolution. 

We study the effects of the cooperator positions in Poisson networks and scale-free networks, which have been widely used in models of cooperation. We use standard versions of these networks that have been commonly used in other studies of cooperation in networks (\cite{Roca2009, Szabo2007}), as well as versions with increased degree assortativity (where individuals of similar connectedness are more likely to be linked to each other in the network; (\cite{Newman2002, noldus2015assortativity})). Such assortativity is frequently observed in real-world networks and is particularly likely to affect the evolution of cooperation when strategy is correlated to degree, because it then affects the extent to which cooperators are connected to each other. 

We study the evolution of cooperation in these networks for two fundamental and commonly used game-theoretical formalisations of cooperative interactions, the Prisoner’s Dilemma game and the Snowdrift game. The Prisoner’s Dilemma game represents a situation where behaving cooperatively in itself is not beneficial to the actor, and cooperation cannot survive in a well-mixed (i.e. unstructured) population (without special mechanisms). This game embodies the paradox of the evolution of cooperation. The Snowdrift game represents a weaker social dilemma where behaving cooperatively in itself provides a benefit to the actor, and a well-mixed population in equilibrium can contain both cooperators and defectors.

\section{Model}

\subsection{General modelling framework}

The model simulates the dynamics of a cooperative strategy in network structures, with interactions between individuals (nodes) occurring across the network edges. While the network structure does not change throughout a simulation, the individuals change their strategies over time, and the main outcome of the simulation is the frequency of cooperators in the population after a set number of timesteps. Each timestep consists of an \textit{interaction phase}, where all individuals connected by a direct edge interact pairwise, and an \textit{update phase}, where all individuals update their strategy adaptively. \vspace{\baselineskip}

We focus on two-player, symmetric games with a binary choice of strategies, as is commonly done in models of cooperation. The game is determined by the following payoff matrix

\begin{equation}
\label{eq.payoffM}
\begin{array}{cc}
\begin{blockarray}{cccl}
& & \text{\scriptsize{cooperate}} & \text{\scriptsize{defect}}  \\
\begin{block}{c r[cc]}
\multirow{2}{*}{$M = $\,\,} & \text{\scriptsize{cooperate}}\,\, &  R & S \\
& \text{\scriptsize{defect}}\,\, &  T & P \\
\end{block}
\end{blockarray}
\end{array}
\end{equation}

For each node $i$ in the network, we will denote the strategy adopted by the corresponding individual by $s_i$. The payoff for individual $i$ when playing against individual $j$ is then $M_{s_i s_j}$. 

Two well-known instances of games of the above form are the Prisoner’s Dilemma game and the Snowdrift game. They both formalize a situation where it is of advantage for the individual to defect (having $T$ as the highest payoff), but if both individuals defect they are worse off than if they both cooperate. In Prisoner’s Dilemma, the worst outcome is to be defected upon while cooperating, with the order of the payoffs being $T>R>P>S$, whereas in the Snowdrift game, the worst is to be defected upon while defecting, with the payoff order being $T>R>S>P$. Note that in Prisoner’s Dilemma, the strategy with the highest individual payoff is to defect regardless of the opponent’s strategy. In well-mixed populations (corresponding to networks where all nodes are connected directly to each other), evolution therefore selects for defection, and cooperation does not survive. In the Snowdrift game, the best payoff depends on the opponent’s strategy, and cooperation and defection can co-exist in unstructured populations. 

In our simulations, we use common one-parameter versions of the two games (\cite{Hauert2004, Nowak1992, Santos2005, Santos2006}), where the severity of the social dilemma (how hard it is for cooperation to evolve, everything else equal) is determined by a single parameter. For Prisoner's Dilemma, we set $R = 1$ and $P=S=0$, and the game is parameterized by the benefit to defectors $b=T$.  For $b=1$ there is no dilemma, while larger values represent larger temptation to defect (making it harder for cooperation to evolve). As is often done, we take $1 \leq b \leq 2$. The Snowdrift game is parameterized by the cost-to-benefit ratio of mutual cooperation $0 < \rho \leq 1$, with $T=\frac{1}{2}(\rho^{-1}+1)$, $R=\frac{1}{2}\rho^{-1}$, $S=\frac{1}{2}(\rho^{-1} - 1)$, and $P=0$. In unstructured populations, $1-\rho$ is the equilibrium fraction of cooperators (for replicator dynamics). \vspace{\baselineskip}

In the interaction phase of each simulation timestep, each individual plays a single game round with each of its network neighbors. We define an individual’s \textit{fitness} in a given timestep to be its summed game payoffs for that timestep. That is, for an individual defined by a node $i$, the fitness is

\begin{equation}
\label{eq.fitness}
F_i = \sum_{j \in \mathcal{N}_i} M_{s_i s_j} ,
\end{equation}
where $\mathcal{N}_i$ is the set of neighbouring nodes of $i$. 

The simulation proceeds to the update phase when all network neighbours have interacted. Here, each individual decides whether to change its strategy, based on how well it did in the interaction phase in terms of fitness. Strategy update is synchronous and follows the \textit{proportional imitation} update rule (\cite{Hauert2004, Santos2005}). For an individual defined by node $i$, a neighbour $j$ is chosen uniformly at random from the set of neighbours $\mathcal{N}_i$. If the neighbour has higher fitness than $i$, that is $F_j > F_i$, then $i$ adopts its strategy with probability 

\begin{equation}
\label{propimit}
\frac{F_{j}-F_{i}}{\max\{k_i,k_j\} D}  ,
\end{equation} 

where $k_i$ denotes the degree of node $i$, and $D$ is the difference between the largest and smallest payoffs for the given game ($D=T-S$ for Prisoner's Dilemma and $D=T-P$ for Snowdrift). The denominator ensures normalisation of the probability. We note that the above update rule corresponds to replicator dynamics adjusted to structured, finite populations (\cite{Hauert2004, Hofbauer1998, Santos2005}), and that it assumes that individuals do not have perfect information on their neighbours (a relevant assumption for many cases in social systems). Also note that the update phase can alternatively be interpreted as reproduction, in which case each timestep is a generation.

\subsection{Networks}

We use four types of networks: standard versions of Poisson and scale-free networks, and versions of these networks with the same degree distributions but with increased degree assortativity. All networks have $N=10^3$ nodes (large enough to avoid boundary effects, \cite{Santos2005}) and an average degree of $\bar{k}=10$, and we use only networks where all nodes are contained in a single component, i.e. all individuals are at least indirectly connected to each other. 

For the standard networks, we use Poisson networks of the Erd\H{o}s-R\'{e}nyi type (\cite{Erdos1960}) and scale-free networks of the Barab\'asi-Albert type (\cite{Barabasi1999}). To generate versions of these networks with increased degree assortativity, we apply the algorithm introduced by Xulvi-Brunet and Sokolov (\cite{Xulvi2004}), which preserves the degree distribution of the network. The algorithm consists of iterated rewiring rounds. In each rewiring round, two edges of the network are chosen uniformly at random and one of two rewiring schemes are carried out: \textit{(i)} with probability $p$ the edges are rewired such that one edge connects the two nodes of highest degree and one connects the two nodes of lowest degree (if this is not already the case); \textit{(ii)} with probability $1-p$ the edges are rewired at random. The degree assortativity of the network can thus be controlled by varying $p$. We use $p=1$ (i.e. maximal degree assortativity given the degree distribution and the condition of all nodes belonging to the same component). The rewiring procedure must be repeated sufficiently many times that almost all edges have been rewired, i.e.~such that every edge has been selected for rewiring with high probability. Denoting the total number of edges in the network by $L$, after $\tau$ iterations the probability that a given edge has \textit{not} yet been selected is $(1-2/L)^\tau \approx e^{-2\tau/L}$ for large $L$. The number of edges not yet selected is thus approximately $L e^{-2\tau/L}$. Requiring this number to be of order unity, we see that we need $\tau \approx L \log(L)/2$ iterations. To make sure we reach maximum assortativity for a given network, we take $\tau = 10 L\log(L)$.

\subsection{Correlations between cooperative strategy and social connectedness}

 We use three levels of correlation between cooperative strategy and social connectedness (degree), which we create by using different methods for how strategies (\textit{cooperate} and \textit{defect}) are assigned to nodes in the beginning of the simulations. Denoting the total number of nodes by $N$ and the number of cooperators by $N_c$, the fraction of cooperators in the population is $r = N_c/N$. The initial fraction of cooperators is $r_{in}$, and we take $r_{in} = 1/2$. To create the three levels of strategy-degree correlations, We use the following strategy assignment procedures:

\begin{itemize}

\item[(1)] \textit{No correlation (uniform assignment)}. Here, cooperators are placed randomly in the network and there is no correlation induced between strategy and connectedness, giving a baseline for our investigations. $N r_{in}$ nodes are picked uniformly at random among all nodes and assigned the cooperator strategy, and the remaining nodes are assigned the defector strategy. The probability for any given node of being a cooperator thus equals $r_{in}$. 

\item[(2)] \textit{Intermediate correlation (stochastic-by-degree assignment)}. Here, cooperators are placed preferentially on high-degree nodes, but with stochasticity in the placement, creating intermediate correlation between strategy and connectedness. Nodes to be assigned the cooperator strategy are selected sequentially based on their relative degree. The first cooperator node is drawn among all nodes, with the probability of drawing node $i$ given by $k_i / \sum_j k_j$, where $k_i$ is the degree of node $i$. Each subsequent cooperator node is drawn from the remaining set of nodes according to $k_i / \sum_{j\notin\mathcal{C}} k_j$, where $\mathcal{C}$ is the set of nodes which have already been selected. This is iterated until $N r_{in}$ nodes have been assigned the cooperator strategy. The remaining nodes are assigned the defector strategy \footnote{Note that this is reminiscent of probability-proportional-to-degree sampling without replacement.}.

\item[(3)] \textit{Perfect correlation (deterministic-by-degree assignment)}. Here, cooperators are placed on the highest-degree nodes without any stochasticity in the placement, giving perfect correlation between strategy and connectedness. The $N r_{in}$ nodes of highest degree are assigned the cooperator strategy and the rest are assigned the defector strategy.

\end{itemize}

\subsection{Simulation procedures}

We run simulations for all combinations of the two games, the four network types, and the three levels of strategy-degree correlation (strategy assignment methods) described above. For each of these 24 combinations, we run simulations for different severities of the social dilemma (that is, for different values of the game parameters $b$ and $\rho$). We run 50 replications for each setting (i.e. for each combination of game, network type, strategy-degree correlation level, and game parameter value). All simulations have a total of $t_{max}=10^4$ timesteps, and the average final fraction $r_{fin}$ of cooperators for a given setting is calculated as the average fraction in the last 100 timesteps of the 50 replications.

\begin{figure}[b]
\begin{center}
\includegraphics[width=\columnwidth]{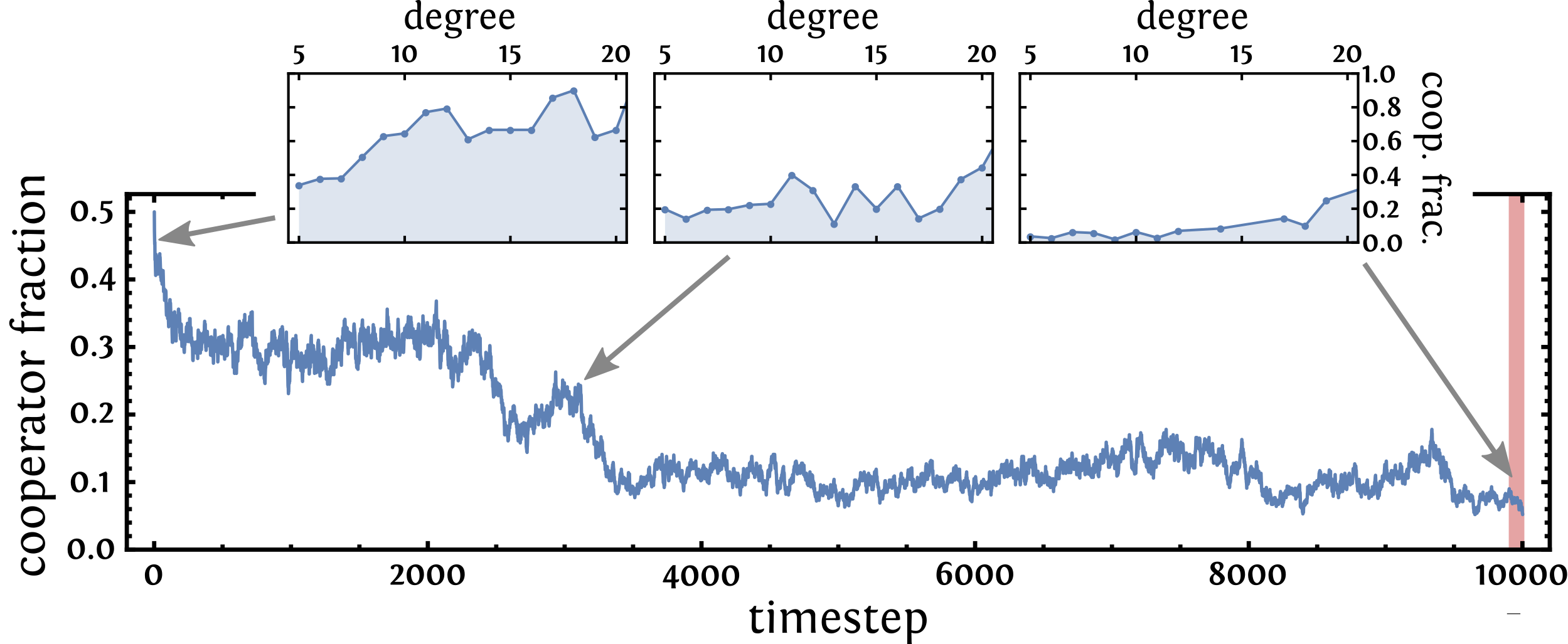}
\caption{An example of a single simulation run of the evolution of cooperation in a network with correlation between strategy and connectedness. The graph shows the fraction of cooperators over time in a standard scale-free (Barabási-Albert) network for Prisoner's Dilemma with  $b=2$ and intermediate correlation between cooperative strategy and degree. The red shaded region indicates the last 100 generations used to compute the final fraction of cooperators $r_{fin}$. \textit{Insets:} Snapshots of the cooperator fraction vs.~node degree at timesteps 1, 3000, and 10000, for degree 1-20 (above 20 there are only few nodes per degree). }
\label{fig.cfracexample}
\end{center}
\end{figure}

An example of a simulation run (a single replication) is shown in Figure~\ref{fig.cfracexample}. In this particular example, the cooperator fraction drops from the initial value of 0.5 to close to zero at the end of the simulation, i.e. cooperation approaches extinction. The example is for Prisoner's Dilemma with intermediate strategy-degree correlation (\textit{stochastic-by-degree} strategy assignment), and the insets indicate that higher-degree nodes, as expected, tend to be more likely to be cooperators.

\begin{figure*}[t]
\begin{center}
\includegraphics[width=\textwidth]{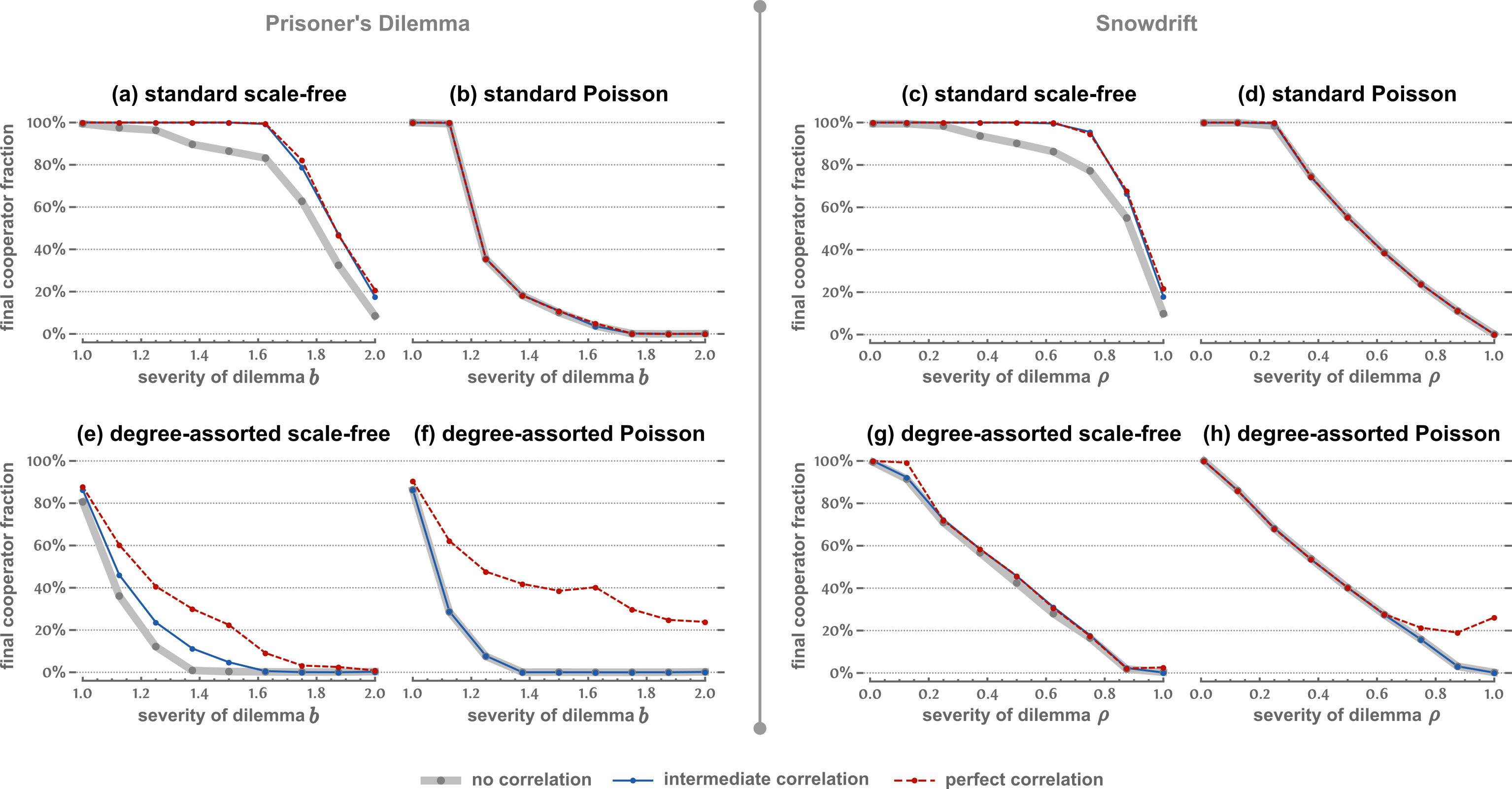}
\caption{Simulation results of the evolution of cooperation in networks with different levels of correlation between cooperative strategy and social connectedness. It can be observed that correlations between strategy and connectedness can increase the success of cooperation, and that the effect depends on the combination of game, network type, and level of strategy-connectedness correlation. Results for the Prisoner’s Dilemma game and the Snowdrift game are to the left and right respectively. In the upper row are shown results for standard scale-free and Poisson networks, and in the lower row are results for versions of these networks with increased degree assortativity. The average final fraction of cooperators is plotted against the severity of the social dilemma (game parameters $b$ and $\rho$). The three curves on each plot are for three different levels of correlation between cooperative strategy and connectedness (see legend).}
\label{fig.cfracres}
\end{center}
\end{figure*}

\section{Results}

The simulation results are shown in Figure~\ref{fig.cfracres}. For the standard networks (Figure~\ref{fig.cfracres}, top row of plots), the social connectedness of cooperators has an effect on the evolution of cooperation only in scale-free networks, for both games. For standard scale-free networks (Figure~\ref{fig.cfracres}a,c), the average final cooperator fraction is increased for the whole game parameter range for both games, when there is correlation between cooperative strategy and degree (intermediate or perfect). For standard Poisson networks (Figure~\ref{fig.cfracres}b,d), the strategy-degree correlations have negligible effect for both games. Also, for each of the standard networks, the final cooperator fractions for the intermediate and perfect correlation between strategy and degree are almost indistinguishable. In the standard networks, the effect of cooperator connectedness thus mainly depends on the network type (scale-free vs. Poisson), with similar results for the two games and the two types of non-zero correlation. The results show that when cooperators occupy well-connected network positions, it increases cooperation in scale-free networks. 

The picture is somewhat different for the networks with increased degree assortativity (Figure~\ref{fig.cfracres}, bottom row of plots). For degree-assorted scale-free networks, both types of strategy-degree correlation (intermediate and perfect) enhance cooperation for Prisoner’s Dilemma, with the perfect correlation having a larger effect (Figure~\ref{fig.cfracres}e). For the Snowdrift game, however, the effects in this network type are very limited (Figure~\ref{fig.cfracres}g). For degree-assorted Poisson networks, the intermediate strategy-degree correlation does not affect cooperation (Figure~\ref{fig.cfracres}f,h). In contrast, the perfect correlation has a large, positive effect for all $b>1$ for Prisoner's Dilemma, and also enhances cooperation for large $\rho$ for the Snowdrift game. Hence, for the networks with increased degree assortativity, the effect of correlation between cooperative strategy and social connectedness depends non-trivially on the combination of the network type, the game, and the level of correlation (presence of stochasticity in the cooperator placement). Here, cooperators’ occupation of well-connected network positions can enhance cooperation in both scale-free and Possion networks, under certain conditions.

\begin{figure*}[t]
\begin{center}
\includegraphics[width=\textwidth]{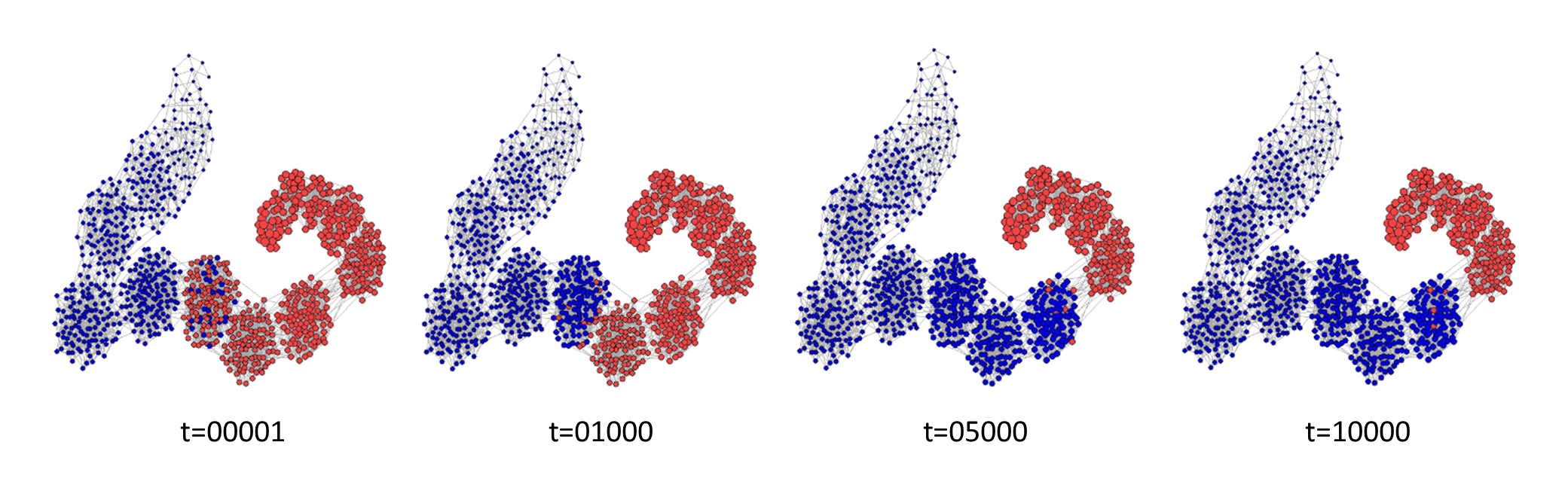}
\caption{Strong correlations between strategy and connectedness in degree-assorted networks can lead to dynamics where the defector strategy is held back at bridge areas between clusters. The figure shows an example of strategy distributions for different timesteps $t$ in a Poisson network with increased degree assortativity and perfect correlation between strategy and degree (for the Prisoner’s Dilemma and \textit{b}=2). Cooperators are shown in red and defectors in blue, and larger node size indicates higher degree.}
\label{fig.networksovertime}
\end{center}
\end{figure*}

\section{Discussion}

Our results support the intuitive expectation that well-connected cooperators can enhance the spread and persistence of cooperation, but also show that they only do so under certain conditions. We see that in standard networks (without increased degree assortativity), the enhancement depends on the network type, with correlation between cooperative strategy and social connectedness (degree) having an effect on final cooperator fractions in scale-free networks and negligible effect in Poisson networks. In degree-assorted networks, we find a more involved pattern of effects, and further investigation of our results confirm that this pattern is due to specific strategy dynamics arising from the modular nature of the degree-assorted networks. Here, bridge areas between clusters of individuals with similar degree can act as barriers for the spread of the defective strategy when cooperators are occupying high-degree nodes, which can lead to a large increase (towards 40 percent) in the average final cooperator fraction; but this increase is sensitive to \textit{Trojan horses}, i.e. defectors placed within the cooperator clusters (when the correlation between strategy and degree is imperfect), which can counteract the protective effect of the bridge areas. In the following we go into further details about these insights. We first consider why well connected cooperators sometimes—but not always—enhance cooperation, and we thereafter focus on the dynamics behind the results in degree-assorted networks. \vspace{\baselineskip}

\subsection{Presence and absence of enhancement of cooperation when cooperators are well connected}

To understand why well-connected cooperators sometimes—but not always—enhance the spread and persistence of cooperation, we may consider the mechanisms by which increased social connectedness can help them promote their strategy. Firstly, the cooperators’ higher number of edges means that they are more likely to be picked as role models when individuals update their strategy (while this is not true in the limit of perfect degree assortativity, we checked that it holds true for all the network types used here). Secondly, well-connected cooperators have higher maximal fitness (because a higher number of social partners means increased opportunity to gain benefits from cooperative interactions, c.f.~Eq.~\eqref{eq.fitness}), and this increases the chance that their strategy will actually be copied when they are used as role models (because individuals only copy strategies from neighbors with higher fitness than themselves). Together, these circumstances can increase the chance that cooperators will spread their strategy to neighbors, when the relative connectedness of cooperators is increased. 

\begin{figure*} [t]
\begin{center}
\includegraphics[width=0.7\textwidth]{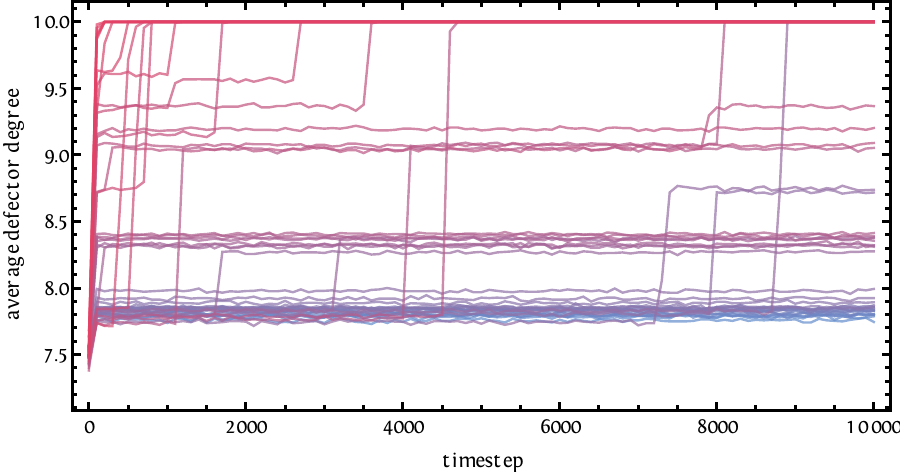}
\caption{Dynamics of defector invasions in networks with increased degree assortativity and strong correlations between cooperative strategy and social connectedness. It can be observed how the average defector degree increases in rapid steps corresponding to the overtaking of network clusters (where many defectors increase their degree within a relatively short time), which are separated by long periods of no change that correspond to the defector strategy being held back at the bridge areas between clusters. The average defector degree over time is shown for 50 simulations (one line for each), for Poisson networks with increased degree assortativity, under the Prisoner’s Dilemma and perfect correlation between cooperative strategy and social connectedness (degree). The average defector degree was measured every 100 time steps.}
\label{fig.Ddegsovertime}
\end{center}
\end{figure*}

\vspace{\baselineskip}

However, we see from the results that correlation between cooperative strategy and social connectedness does not always enhance cooperation (even under perfect correlation). This has to do with the fact that cooperators and defectors do not get the same payoff from the same neighborhood, meaning that everything else equal, a cooperator having a higher degree than a defector is not always enough for it to gain higher fitness than the defector. More specifically, a cooperator generally needs to have considerably more cooperator neighbors than a defector, to get a higher fitness than the defector (see Appendix A for details). A higher degree for the cooperator will not always lead to this requirement being satisfied, and therefore does not necessarily give the cooperator advantage over the defector. In the standard networks, we saw positive effects of the correlations between cooperativeness and degree in scale-free networks, whereas there were no effects in the Poisson networks - even under perfect correlation. Our results thus show that the cooperators’ benefit from higher social connectedness is too small to affect the final outcome in standard Poisson networks, whereas in standard scale-free networks the advantage cooperators gain from the increased connectedness leads to positive effects on the evolution of cooperation.
\vfill

\subsection{Degree assortativity creates barriers that can strongly enhance cooperation, but their effect can be counteracted by Trojan horses}

\begin{figure*} [t]
\begin{center}
\includegraphics[width=0.9\textwidth]{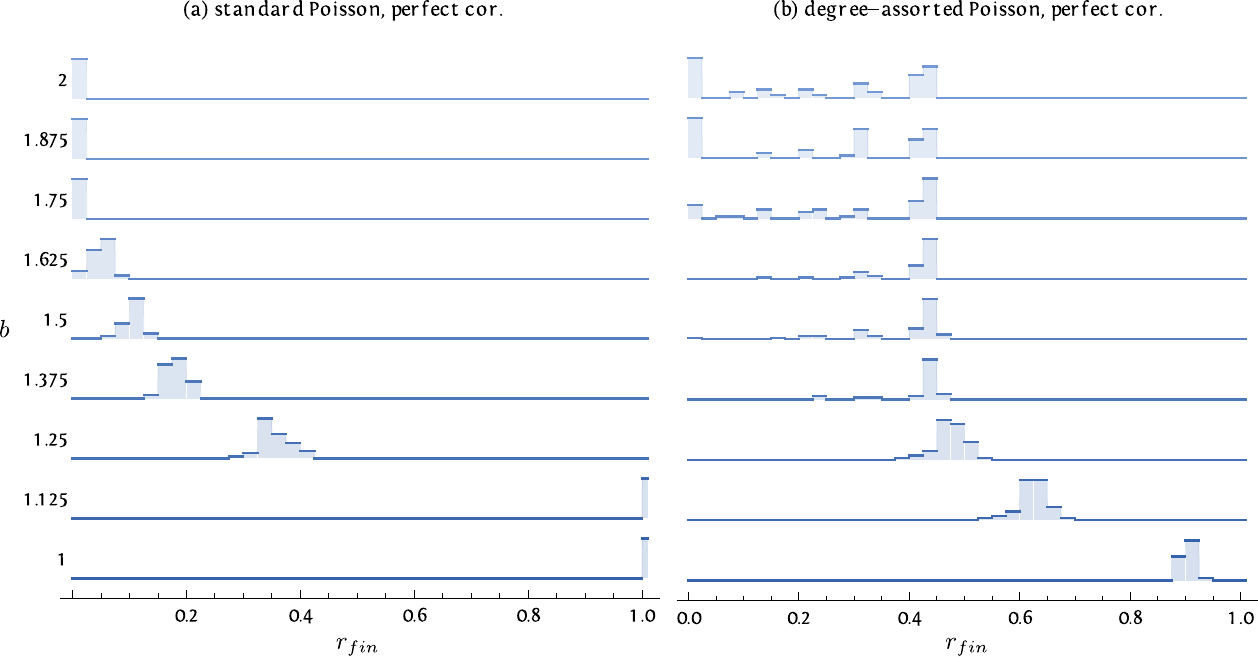}
\caption{Distributions of final cooperator fractions for Poisson networks with strong correlations between cooperative strategy and social connectedness, with and without increased degree assortativity. It can be observed how the final cooperator fractions occur in distinct peaks for the degree-assorted network, due to the defector strategy being held back at bridge areas between clusters. The final cooperator fraction $r_{fin}$ is shown for different values of the game parameter $b$. The results are for the Prisoner’s Dilemma with perfect correlation between cooperative strategy and social connectedness (degree), for \textbf{(a)} standard Poisson networks, and \textbf{(b)} Poisson networks with increased degree assortativity. Each distribution is based on 50 replicates and scaled to its maximal value.}
\label{fig.finalcfracdists}
\end{center}
\end{figure*}

In the networks with increased degree assortativity, we see an interesting pattern with strong enhancements of cooperation under the perfect correlation between cooperative strategy and social connectedness, in particular in Poisson networks where the average final cooperator fraction stays relatively high throughout the game parameter ranges (above about 20 percent, Figure~\ref{fig.cfracres}). This enhancement comes about because of the structure of the degree-assorted networks, where nodes are clustered by degree. When cooperators occupy the high-degree nodes, the bridge areas between clusters act as barriers against the defector strategy, leading to dynamics where the defector strategy invades a cluster rapidly and then is stopped by the bridges, where it then waits until it might eventually overcome them and invade the next cluster. This limits the spread of defection and thereby leads to the high final cooperator fractions. These dynamics can be seen from further investigation of the simulation results. An example of how the defector strategy is stopped at bridge areas for a single simulation is shown in Figure~\ref{fig.networksovertime}. More generally, the pattern is evidenced by the average defector degree over time, which increases in very rapid steps (corresponding to the overtaking of clusters), separated by long periods of no change (corresponding to the spread of defection being hindered by bridge areas; Figure~\ref{fig.Ddegsovertime}). And finally, the dynamics are also reflected in the distributions of final cooperator fractions (Figure~\ref{fig.finalcfracdists}), where we see that in the degree-assorted network, these fractions occur in multiple distinct peaks that correspond to the defector strategy waiting by the various bridge areas (a pattern that arises because the simulation is statistically unlikely to stop during the rapid cluster invasions). 

We do not observe a strong enhancement of cooperation in the degree assorted networks when the correlation between cooperative strategy and degree is intermediate. In this case, the protection from the bridge areas is counteracted by what we term \textit{Trojan horses}, i.e. defectors placed within cooperator-dominated clusters, which facilitate the spread of the defective strategy from within the clusters. Our results thus show that although the extent of the strategy-degree correlations (intermediate or perfect) made virtually no difference in the standard networks, it can make a major difference under increased degree assortativity, via the Trojan-horse effect. 

While this study focuses on the effect of correlations between strategy and connectedness on cooperation, the results also add to the knowledge about the effect of degree assortativity. Degree assortativity (also known as degree mixing or degree degree correlation) has been studied in various settings, including multilayer networks (e.g. \cite{wang2014degree, duh2019assortativity}) and evolving networks (e.g. \cite{tanimoto2010effect}), and analytical measures for the heterogeneity of networks with degree assortativity have been developed (\cite{devlin2009cooperation}). The effect of degree in the absence of any degree variation (i.e. perfect degree assortativity) has also been investigated (e.g. \cite{stella2022lower}). Rong et al. studied the effect of directly tuning the degree assortativity in static networks, similarly to the method we used here to create versions of standard networks with increased degree assortativity (\cite{Rong2007}). They found that increased degree assortativity in itself can decrease cooperation, which is also seen in our results under no correlation and intermediate correlation between cooperative strategy and social connectedness. Interestingly, we observe that the perfect correlation between strategy and connectedness counteracts the negative effect of degree assortativity to such an extent that in the Poisson networks, cooperation does better in the degree-assorted networks than in the standard networks (Figure~\ref{fig.cfracres}f,h vs. Figure~\ref{fig.cfracres}b,d). \vspace{\baselineskip}

We also note that under certain circumstances (the Prisoner’s Dilemma with perfect strategy-degree correlation and degree assortativity; Figure~\ref{fig.cfracres}e,f), cooperation does better under Poisson degree distribution than under scale-free degree distribution. This is opposite to the pattern found for standard Poisson and scale-free networks (Figure~\ref{fig.cfracres}a,b and previous studies, e.g. \cite{Santos2006}), where cooperation does much better under scale-free degree distribution than under Poisson distribution. This demonstrates that the generally observed positive relationship between degree heterogeneity and cooperation (\cite{Santos2006evol, Santos2012}) is under some conditions overruled. \vspace{\baselineskip}
\vfill

\subsection{Further remarks}

Our study implies that correlation between cooperativeness and social connectedness is a relevant factor to consider both for our understanding of the evolution of cooperation in general and when designing and conducting scientific investigations of cooperation. Empirical experiments of humans playing cooperation games in constructed networks constitutes an important tool in the study of cooperation 
(e.g. \cite{Cassar2007, Gracia2012, grujic2020people, li2018punishment, melamed2018cooperation, Rand2014}, see also \cite{jusup2022social}). Given that the number of replicates in such experiments can be low due to practical constraints, random variation in initial conditions can have significant effects. Our results imply that if cooperative individuals are by chance placed in more well-connected network positions, this can potentially affect the experimental results. We also saw that the amount of variation in the final cooperator fractions differed between settings, with some of them having more propensity for outliers than others (Figure~\ref{fig.finalcfracdists}, and see Appendix B for similar plots for the other settings), increasing the chance of getting unrepresentative experimental results. Such effects can therefore be important to consider when conducting network cooperation experiments. Similarly, the results show that initial conditions (here in the form of strategy-degree correlations) can have considerable effects in simulation studies of cooperation in networks, which underlines the importance of using replications in such studies. 

We have looked at the evolution of cooperation in static networks (where the structure does not change). This allowed us to investigate how the effect of correlations between cooperative strategy and social connectedness depends on the network structure. While social systems of humans and other species are dynamic in the sense that they consist of series of time-limited social interactions, temporal changes in the emerging social network structures (that is, termination and emergence of social bonds) may be slow, which is supported by the finding of long-term stability in the structure of social networks in multiple species (\cite{borgeaud2017,godfrey2013,kerth2011,prehn2019}). When changes in the real social structures are much slower than the rate by which individuals change their cooperative strategies, then static social networks can approximate the systems well in investigations of the evolution of cooperation. Our results imply that correlations between cooperative strategy and social connectedness can potentially contribute to the persistence and spread of cooperation in real-world networks with stable social structures. Such effects may also be relevant in unstable networks, depending on the nature of the social linking dynamics (\cite{santos2006coop,zimmermann2005, tanimoto2010effect, duh2020mixing}). Investigations of the role of correlations between strategy and network position for the evolution of cooperation in real and simulated social systems with different levels of stability (\cite{li2020evolution}) constitute an exiting avenue for future research.

\subsection*{Acknowledgements}
This research was supported by a grant from the Carlsberg Foundation (grant number: CF20-0663) to Josefine Bohr Brask, and a grant from the Independent Research Fund Denmark (grant number: 7027-00044B) to Jonatan Bohr Brask. We thank Sylvia Dimitriadou, Tim Fawcett, and Andrew Higginson for helpful comments on an earlier version of the manuscript.

\bibliography{Final_preprint_BraskBrask_2024}
\clearpage
\onecolumngrid

\section*{Appendix A: Neighbourhoods and fitness}

Consider a cooperator node of degree $k$ with $n$ cooperator neighbours and a defector node of degree $k'$ with $n'$ cooperator neighbours. The fitness of two such nodes for games with payoffs such as those used here (formalised in Eq.~\eqref{eq.payoffM})  will be respectively
\begin{equation}
F = n R + (k-n) S ,
\end{equation}
and
\begin{equation}
F' = n' T + (k'-n') P .
\end{equation}
The cooperator has higher fitness than the defector when $F>F'$. For the one-parameter Prisoner's Dilemma game, the payoffs are $S=P=0$, $R=1$, and $T=b$, and thus $F>F'$ if and only if
\begin{equation}
n > b n' .
\end{equation}
That is, a cooperator must have $b$ times as many cooperator neighbours as a defector to gain a higher fitness. For the one-parameter Snowdrift game, we have $T=\frac{1}{2}(\rho^{-1}+1)$, $R=\frac{1}{2}\rho^{-1}$, $S=\frac{1}{2}(\rho^{-1} - 1)$, and $P=0$. In this case $F>F'$ when
\begin{equation}
n \rho + k(1-\rho) > n' (1+\rho) .
\end{equation}
We see that for the Snowdrift game, the cooperator also benefits from a high number of cooperator neighbours, but a sufficiently high degree can compensate for a low number of cooperator neighbours. The inequality is always fulfilled for $k=n'(1+\rho)/(1-\rho)$. However, for cost-to-benefit ratios approaching 1 this diverges, and so in this regime $n$ must again be larger than $n'$. For the most severe instances of both games, i.e.~for the parameter settings making it hardest for cooperation to evolve ($b=2$ and $\rho=1$), cooperators need to have more than twice as many cooperator neighbours as defectors to achieve higher fitness.

\section*{Appendix B: Final cooperator fraction figures}

Distributions of final cooperator fractions for the Prisoner’s Dilemma game and the Snowdrift game, for all combinations of the four network types and the three levels of strategy-degree correlation. For each combination, the distribution is shown for each used value of the respective game parameter.

\begin{figure*} [b!]
\begin{center} 
\includegraphics[width=0.9\textwidth]{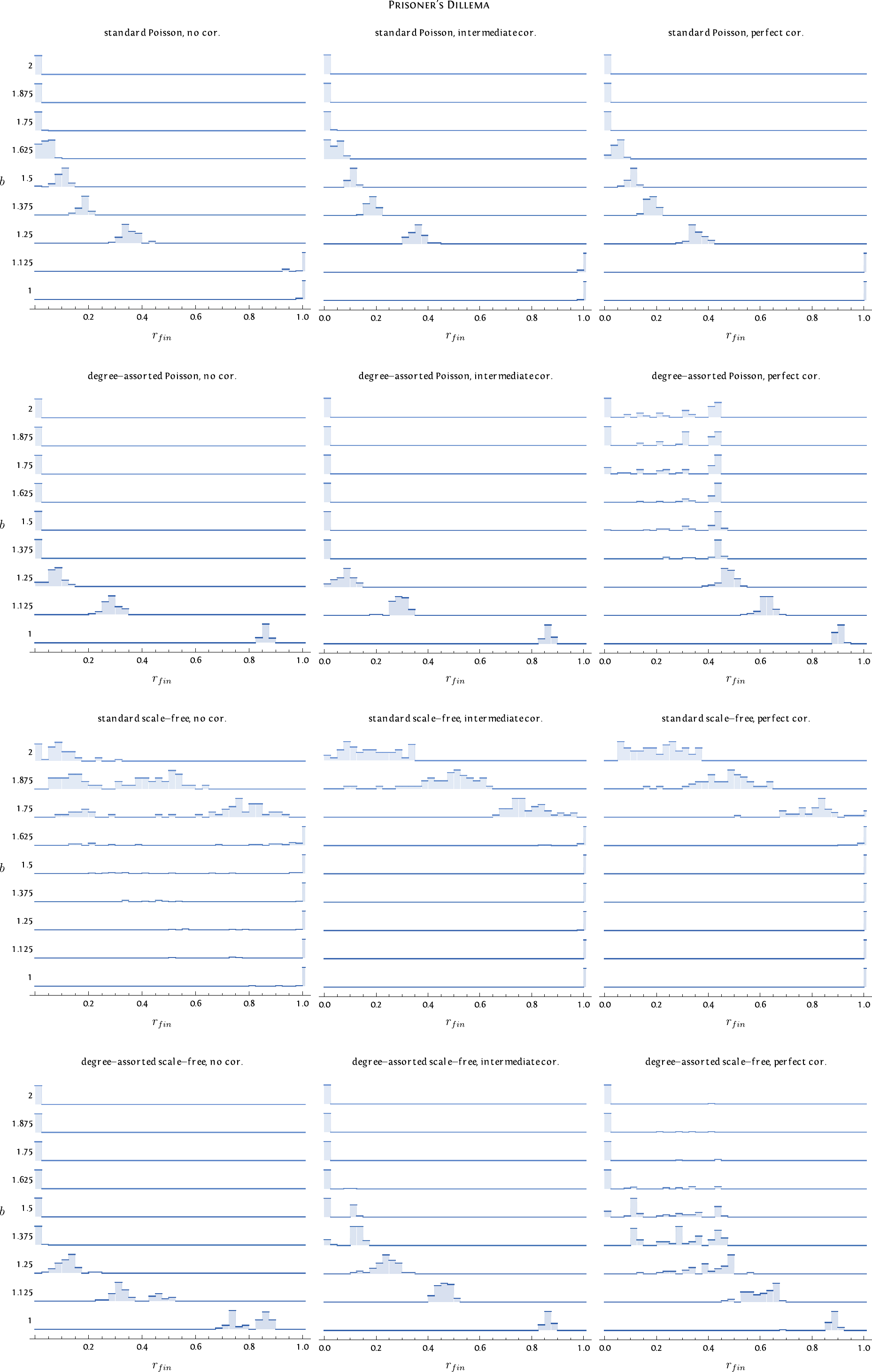}
\label{fig.suppl1}
\caption{Distributions of $r_{fin}$ for Prisoner's Dilemma.}
\end{center}
\end{figure*}

\begin{figure*} 
\begin{center}
\includegraphics[width=0.9\textwidth]{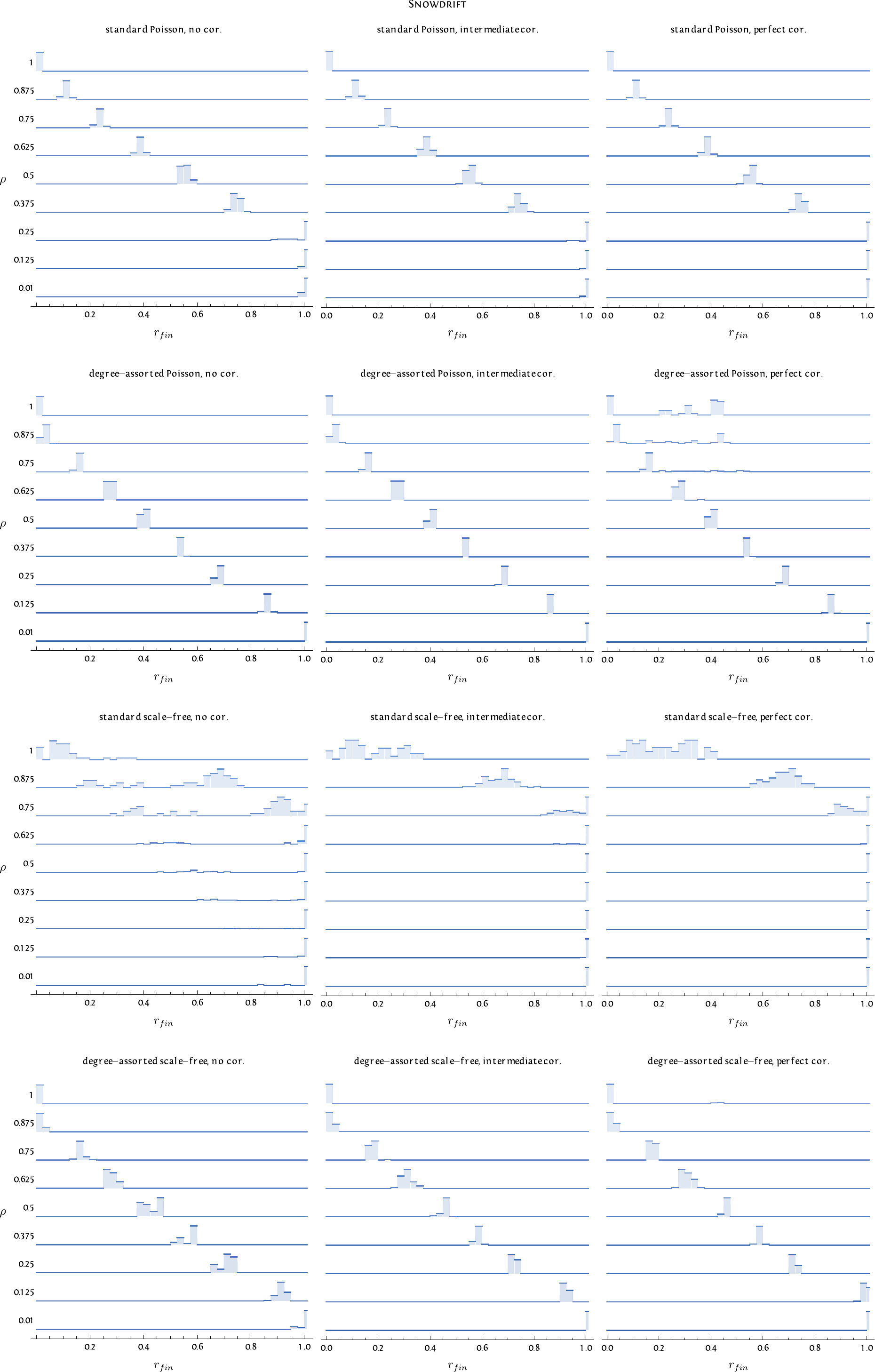}
\label{fig.suppl2}
\caption{Distributions of $r_{fin}$ for Snowdrift.}
\end{center}
\end{figure*}

\end{document}